%Paper: hep-th/9403046
%From: ereidell@marie.mit.edu (Evan Reidell)
%Date: Mon, 7 Mar 1994 16:40:22 -0500
%Date (revised): Mon, 7 Mar 1994 16:46:24 -0500
%Date (revised): Mon, 7 Mar 1994 16:49:36 -0500

%%%
%%% If anyone has trouble printing the figures on pages 2 and 3,
%%% please send me an email message containing the error messages!
%%% -- ereidell@marie.mit.edu
%%%

\magnification 1200
\font\bfmone=cmbx10 scaled\magstep1
\input epsf
\hsize=6.5truein
\vsize=9.0truein
\pretolerance 700
\tolerance 3000
\overfullrule 0pt

\baselineskip=17.2truept plus .5truept minus .5truept

\def\pmb#1{\setbox0=\hbox{$#1$}%
\copy0\kern-\wd0
\kern-.025em\copy0\kern-\wd0
\kern+.025em\raise.025ex\copy0\kern-\wd0
\kern+.025em\box0}

\def\footnoterule{\kern-3pt \hrule width\hsize \kern2.6pt}
\footline={\ifnum\pageno>1\hfil--\folio--\hfil\else\hfil\fi}

\centerline{\bfmone Topological Structures in the Standard Model at High $T$}
\medskip
\centerline{R.~Jackiw\footnote{*}
{\tenrm This work is supported in part
by funds provided by the U.~S.~Department of Energy
(D.O.E.) under contract \#DE-AC02-76ER03069.}}
\vskip4pt
\centerline{\it Center for Theoretical Physics,}
\centerline{\it Department of Physics, and}
\centerline{\it Laboratory for Nuclear Science,}
\centerline{\it Massachusetts Institute of Technology,}
\centerline{\it 77 Massachusetts Avenue,}
\centerline{\it Cambridge, MA ~02139-4307}
\smallskip
\centerline{\tenrm
(FubiniFest, 24-26 February 1994, Torino ITALY. ~~~ MIT-CTP-2285.)}

\bigskip

We meet to wish our dear colleague Sergio Fubini all the best on his
sixty-fifth birthday.  Moreover, behind the formality of this calendrically
significant instant,
there is also our timeless
expression of great affection and thanks
to Sergio: affection for his sympathetic character and thanks for his
activities in our profession, both within the scientific framework and in the
broader social context.
But for those of us from MIT there is a special feeling
of gratitude towards Sergio, and that is because in the late sixties and early
seventies he was with us and helped shape what can now be seen as the most
recent golden age of physics, thereby establishing at MIT a tradition that
still flourishes today.  You have to appreciate the moment: Steven Weinberg was
on the faculty completing the standard model, while Garbriele Veneziano and
Sergio were inventing what proved to be the physics of the future --- string
theory.  These people have since departed from our University, but Sergio has
kept the legacy vital by visiting us
--- not as frequently as we would have liked --- and
by encouraging continuing contact with his wonderful and talented Italian
compatriot
physicists.  The people who came from Italy to enrich our
department are too numerous to list, but theirs is an ongoing presence,
formalized recently by an agreement with the INFN, and the well-spring of all
this good fortune is Sergio Fubini, whom we all thank.

Sergio is now gone from MIT, but I am certain he wants to be informed of
activity there, so I shall describe  one project, with the hope that it meets
his criteria of simplicity and symmetry.

These days, as high energy particle colliders become unavailable for testing
speculative theoretical ideas, physicists are looking to other environments
that may provide extreme conditions where theory confronts physical reality.
One such circumstance may arise at high temperature $T$, which perhaps can be
attained in heavy ion collisions or in astrophysical settings.  It is natural
therefore to examine the high-temperature behavior of the standard model, and
here I shall report on recent progress in constructing the high-$T$ limit of
QCD.

In studying a field theory at finite temperature, the simplest approach is the
so-called imaginary-time formalism.  We continue time to the imaginary
interval $[0,1/iT]$ and consider bosonic (fermionic) fields to be periodic
(anti-periodic) on that interval.  Perturbative calculations are performed by
Feynman rules as at zero temperature, except that in the conjugate
energy-momentum, Fourier-transformed space, the energy variable $p^0$
(conjugate to the
periodic time variable) becomes discrete --- it is $2\pi n T$,
($n$ integer) for bosons.  From this one immediately sees that at high
temperature --- in the limiting case, at infinite temperature --- the
time-direction disappears, because the temporal interval shrinks to zero.
Also only zero-energy processes survive, since ``non-vanishing energy''
necessarily means high energy owing to the discreteness of the energy variable
$p^0 \sim 2\pi n T$, and therefore
all modes with $n \neq 0$ decouple at large $T$.
In this way a Euclidean three-dimensional field theory becomes
effective at high temperatures and describes essentially static processes.

While all this is quick and simple, it may be physically inadequate.
First of all,
frequently one is interested in non-static processes in real time, so
complicated analytic continuation from imaginary time needs to be made.  Also
one may wish to study amplitudes where the real external energy is neither
large nor zero, even though virtual internal energies are high.

Large $T$
Feynman graphs with external legs carrying limited amounts of energy
and internal lines
characterized by large momenta because $T$ is large
have been dubbed ``hard thermal loops.''
In fact they are a very important feature of high temperature QCD,
because they necessarily arise in a resummed perturbative expansion.$^{1,2}$

Here is the argument.  Consider a one-loop amplitude $\Pi_1(p),$
$$
\Pi_1(p) \equiv \int dk ~ I_1 (p,k) ~~,
$$
given by the graph in the figure.
$$
%%
%% they warn us against doing this, p.6 of dvips manual
%%

\eqalign{
\Pi_1(p) &=~
\vcenter{\hbox to 125pt{\vbox to 52.8pt{\vss\special{" fig1a}}\hss}}
\cr
&\equiv ~~~~~~~~ \, \int dk ~ I_1(p,k)}
{\hskip.5in}
$$
Compare this to a two-loop amplitude
$\Pi_2(p),$
$$
\Pi_2(p) \equiv \int dk ~ I_2 (p,k) ~~,
$$
in which $\Pi_1$ is an insertion, as in the figure below.
$$
\eqalign{
\Pi_2(p) &=~
\vcenter{\hbox to 125pt{\vbox to 52.8pt{\vss\special{" fig1b}}\hss}}
\cr
& \equiv ~~~~~~~~ \, \int dk ~ I_2 (p,k)}
{\hskip.5in}
$$
Following Pisarski,$^1$
I estimate the relative importance of $\Pi_2$ to $\Pi_1$
by the ratio of their integrands,
$$
{\Pi_2 \over \Pi_1} \sim {I_2 \over I_1} = g^2 \, {\Pi_1(k) \over k^2} ~~,
$$
Here $g$ is the coupling constant, and the $k^2$ in the denominator
reflects that we are considering a massless particle as in QCD.  Clearly the
$k^2 \to 0$ limit is relevant
to the question whether the higher order graph can be neglected relative to
the lower order one.
Because one finds that for small $k$ and large
$T$, $\Pi_1(k)$ behaves as $T^2$, the ratio $\Pi_2/\Pi_1$ is $g^2 T^2 / k^2$.
Thus when $k$ is $O(gT)$ or smaller the two-loop amplitude is not negligible
compared to the one-loop amplitude.  Thus graphs with ``soft'' external
momenta [$O(gT)$ or smaller] have to be
included as insertions in higher order calculations.

These so-called ``hard thermal loops,'' {\it i.e.\/} the high-temperature
limits of real-time Feynman graphs with finite external momenta, have become
the object of much study, which culminated with the discovery (Braaten,
Pisarski, Frenkel, Taylor)$^2$
of a remarkable simplicity in their structure.
Specifically, the generating functional for hard thermal loops with only
external gauge field legs,
%% $\Gamma_{\rm HTL}(A)$,
in an $SU(N)$ gauge theory containing $N_F$ fermion
species of the fundamental representation is found (i) to be proportional to
$(N+{1\over2} N_F)$, (ii) to behave as $T^2$ at high temperature, and (iii)
to be gauge invariant.
\def\myfig#1{%%
\vcenter{\hbox to 52.8pt{\vbox to 54.9pt{\vss\special{" #1}}\hss}}}
$$
\eqalign{
\myfig{2 0 doit}~+\myfig{3 -90 doit}+\myfig{4 45 doit}+~\myfig{5 90 doit}
{}~+ \cdots &= (N + {\textstyle{1\over2}} N_F) \, {g^2 T^2 \over 12\pi} \,
\Gamma_{\rm HTL} (A) \cr
\Gamma_{\rm HTL} (U^{-1} \, A \, U + U^{-1} \, dU) &=
\Gamma_{\rm HTL} (A) }
$$
A further kinematical simplification in
$\Gamma_{\rm HTL}$ has also been established.
To explain this we define two light-like four-vectors
$Q^\mu_{\pm}$
depending on a unit three-vector $\hat{q}$,
pointing in an arbitrary direction.
$$
\eqalign{
Q^\mu_\pm &= {1\over\sqrt{2}} (1,\,\pm \hat{q}) \cr
\hat{q} \cdot \hat{q}  = 1 ~~,~~~~
Q^\mu_\pm Q_{\pm \mu} &= 0 ~~,~~~~
Q^\mu_\pm Q_{\mp \mu}  = 1}
$$
Coordinates and potentials are projected onto $Q_\pm^\mu$
$$
x^\pm \equiv x_\mu Q_\pm^\mu ~~,~~~~
\partial_\pm \equiv Q_\pm^\mu {\partial \over \partial x^\mu} ~~,~~~~
A_\pm \equiv A_\mu Q_\pm^\mu
$$
The additional fact that is now known is that (iv)
after separating an ultralocal contribution from
$\Gamma_{\rm HTL}$, the remainder may be written as an average over
the angles of
$\hat{q}$
of a functional $W$ that  depends only on $A_+$;
also this functional
is non-local only on the
two-dimensional $x_\pm$ plane, and is ultralocal in the remaining directions,
perpendicular to the $x_\pm$ plane.  [``Ultralocal'' means that any potentially
non-local kernel $k(x,y)$ is in fact a $\delta$-function of the difference
$k(x,y) = \delta (x-y)$.]
$$
\Gamma_{\rm HTL} (A) = 2\pi \int d^4x ~ A^a_0 (x) A^a_0(x) +
\int d\Omega_{\hat{q}} \, W (A_+)
$$
These results are established in perturbation theory, and a perturbative
expansion of $W(A_+)$, {\it i.e.\/} a power series in $A_+$, exhibits the
above mentioned properties.  A natural question is whether one can sum the
series, {\it i.e.\/} obtain an expression for $W(A_+)$.

Important progress on this problem was made when it was observed
(Taylor, Wong)$^3$
that the gauge-invariance condition
can be imposed infinitesimally, whereupon it leads
to a functional differential equation for $W(A_+)$, which is best presented as
$$
\eqalign{
& {\partial \over \partial x^+} \, {\delta \over \delta A^a_+}
\left[
W(A_+) + {\textstyle {1\over2}} \int d^4 x ~ A_+^b(x) A_+^b(x) \right]
\cr
& {\hskip.5in} -
{\partial \over -\partial x^-} \left[ A_+^a \right]
+ f^{abc} A_+^b
{\delta \over \delta A_+^c}
\left[
W(A_+) + {\textstyle {1\over2}} \int d^4 x ~ A_+^d(x) A_+^d(x) \right]
= 0}
$$
In other words we seek a quantity, call it
$$
S(A_+) \equiv W(A_+) +{1\over2} \int
d^4 x \, A_+^a (x) A_+^a(x) ~~, $$
which is a functional on a two-dimensional
manifold $\left\{ x^+, x^- \right\}$, depends on a single functional variable
$A_+$, and satisfies
$$
\partial_1 {\delta \over \delta A_1^a} S - \partial_2 A_1^a +
f^{abc} A_1^b {\delta \over \delta A_1^c} S = 0
$$
$$
{\rm ``1\hbox{''}} \equiv x^+ ~~,~~~~
{\rm ``2\hbox{''}} \equiv -x^- ~~,~~~~
A_1^a \equiv A_+^a
$$
Another suggestive version of the above is gotten by defining $A_2^a \equiv
{\delta S \over \delta A_1^a}$.   Then we need to solve
$$
\partial_1 A_2^a - \partial_2 A_1^a + f^{abc} A_1^b A_2^c = 0 ~~.
$$

To solve the functional equation and produce an expression for $W(A_+)$, we
now turn to a completely different corner of physics,
and that is Chern-Simons theory.

The Chern-Simons term is a peculiar
gauge theoretic topological
structure that  can be constructed in odd
dimensions, and here we consider it in  3-dimensional space-time.
$$
I_{\rm CS} \propto \int d^3 x ~ \epsilon^{\alpha \beta \gamma} \,
{\rm Tr} \,
(\partial_\alpha A_\beta A_\gamma + {\textstyle {2\over3}} A_\alpha A_\beta
A_\gamma)
$$
This object was introduced into physics over a decade ago, and since that time
it has been put to various physical and mathematical uses.
Indeed
one of our originally stated motivations
for studying the Chern-Simons term
was its possible relevance
to high-temperature gauge theory.$^4$~
Here following Efraty and Nair,$^5$
we shall employ the Chern-Simons term
for a determination of the hard
thermal loop generating functional, $\Gamma_{\rm HTL}$.

Since it is the space-time integral of a density, $I_{\rm CS}$ may be viewed
as the action for a quantum field theory
in (2+1) dimensional space-time,
and the corresponding Lagrangian
would then be given by a two-dimensional, spatial integral
of a Lagrange density.
$$
\eqalign{
I_{\rm CS} &\propto \int dt ~ L_{\rm CS} \cr
L_{\rm CS} &\propto \int d^2 x ~ \left(
A_2^a \skew{4}\dot{A}_1^a + A_0^a F_{12}^a \right) }
$$
I have separated the temporal index (0) from the two spatial ones (1,2) and
have indicated time differentiation
by an over dot.
$F_{12}^a$ is the non-Abelian field strength, defined on a two-dimensional
plane.
$$
F_{12}^a = \partial_1 A_2^a - \partial_2 A_1^a + f^{abc} A_1^b A_2^c
$$
Examining the Lagrangian, we see that it has the form
$$
L \sim p \dot{q} - \lambda \, H(p,q)
$$
where $A_2^a$ plays the role of $p$, $A_1^a$ that of $q$, $F_{12}$
is like a Hamiltonian and $A_0$ is like the
Lagrange multiplier $\lambda$, which forces the Hamiltonian
to vanish; here $A_0^a$ enforces the vanishing of $F_{12}^{\,a}$.
$$
F_{12}^{\,a} = 0
$$
The analogy instructs us how the Chern-Simons
theory should be quantized.

We postulate equal-time  commutation relations, like those between $p$ and $q$.
$$
\left[ A_1^a ({\bf r}), \, A_2^b ({\bf r}') \right]
= i \, \delta^{ab} \delta({\bf r} - {\bf r}')
$$
In order to satisfy the condition enforced by the Lagrange multiplier, we
demand that $F_{12}^a$, operating on ``allowed'' states, annihilate them.
$$
F_{12}^a | ~~ \rangle = 0
$$

This equation can be explicitly presented in a Schr\"odinger-like
representation
for
the Chern-Simons quantum field theory, where the state is a funtional of
$A_1^a$.  The action of the operators $A_1^a$ and $A_2^a$ is by multiplication
and functional differentiation, respectively.
$$
\eqalign{
\phantom{A_0^a} \, | ~~ \rangle &\sim \Psi(A_1^a) \cr
A_1^a \, | ~~ \rangle &\sim A_1^a \, \Psi(A_1^a) \cr
A_2^a \, | ~~ \rangle &\sim {1\over i} {\delta \over \delta A_1^a}
\, \Psi(A_1^a) \cr
}
$$
This of course is just the field theoretic analog of the quantum mechanical
situation where states are functions of $q$, the $q$ operator acts by
multiplication, and the $p$ operator by differentiation.
In the Schr\"odinger representation,
the condition that states be annihilated by $F_{12}^a$
$$
\left( \partial_1 A_2^{a} - \partial_2  A_1^a + f_{abc} A_1^b
A_2^c \right) \, \Big| ~~ \Big\rangle = 0
$$
leads to a functional differential equation.
$$
\left(
\partial_1 {1\over i} {\delta \over \delta A_1^a}
- \partial_2 \, A_1^a
+ f_{abc} A_1^b \, {1\over i} \, {\delta \over \delta A_2^c}
\right)
\Psi(A_1^a) = 0
$$
If we define $S$ by $\Psi = e^{iS}$ we get equivalently
$$
\partial_1 {\delta \over \delta A_1^a} S - \partial_2 A_1^a + f_{abc} A_1^b
{\delta \over \delta A_2^c} S = 0
$$
This equation comprises the entire content of Chern-Simons quantum field
theory.
$S$ is the Chern-Simons eikonal, which gives the exact wave functional owing
to the simple dynamics of the theory.
Also the above eikonal equation is
recognized to be precisely the equation
for the hard thermal loop generating functional.

The gained advantage is that ``acceptable'' Chern-Simons states,
{\it i.e.\/} solutions to the above functional equations,
had been constructed long ago,$^6$
and one can now take over those results to the hard
thermal loop problem.  One knows
from the Chern-Simons work that $\Psi$ and $S$ are given by
a 2-dimensional fermionic determinant, {\it i.e.\/} by the Polyakov-Wiegman
expression.  While
these are not described
by very explicit formulas, many properties are understood,
and the hope is that one can use these properties to obtain further
information about high-temperature QCD processes.

For example one can compute the induced current
$j^\mu_{\rm induced} \sim {\delta \Gamma_{\rm HTL}
\over \delta A_\mu^a}$, and use this as a source in the Yang-Mills equation,
thereby obtaining a non-Abelian generalization of the Kubo equation, which
governs the response of the hot gluonic plasma to external disturbances.$^7$
$$\eqalign{
D_\mu F^{\mu\nu} &=
{m^2 \over 2} j^\nu_{\rm induced} \cr
m &= gT \sqrt{{N + N_F / 2 \over 3}}
}$$
{}From the known properties   of the fermionic determinant --- hard thermal
loop
generating functional --- one can show that $j^\mu_{\rm induced}$ is given by
$$
j^\mu_{\rm induced} = \int {d \Omega_{\hat{q}} \over 4\pi} \, \left\{
Q_+^\mu \left( \vphantom{1\over1} a_-(x) - A_-(x) \right)
+ Q_-^\mu \left( \vphantom{1\over1} a_+ (x) - A_+(x) \right) \right\}
$$
where $a_{\pm}$ are solutions to the equations
$$
\eqalign{
\partial_+ a_- - \partial_- A_+ + [A_+, a_-] &= 0 \cr
\partial_+ A_- - \partial_- a_+ + [a_+, A_-] &= 0
}$$
Evidently $j^\mu_{\rm induced}$, as determined by the above equations, is a
non-local and non-linear functional of the vector potential $A_\mu$.

An alternative, equivalent derivation of the induced current has been given
by Blaizot and Iancu,$^8$
directly from the QCD field equations.
Their argument may be succintly put in the language of the composite effective
action$^9$ and makes use of two   approximations.     The composite action  is
truncated at the one loop (semi-classical) level --- two-particle irreducible
graphs are omitted.   This comprises the first, dynamical approximation.
Then, in the second, kinematical approximation, the
stationary conditions on the one-loop
action are
shown to lead to
the gauge invariance equation for $\Gamma_{\rm HTL}$.

In the Abelian case, everything commutes and linearizes.
One can determine $a_\pm$ in terms of $A_\mp$.
$$
a_\pm = {\partial_\pm \over \partial_\mp} \, A_\mp
$$
Incidentally, this formula exemplifies the kinematical simplicity,
mentioned above, of hard thermal loops:
the nonlocality of $1/\partial_\pm$ is entirely in
$\left\{ x^+, x^- \right\}$
plane.   With the above form for $a_\pm$ inserted into the
Kubo equation, the solution can be constructed explicitly.
It coincides with the results obtained by Silin long ago,
on the basis of the Boltzmann-Vlasov equation,$^{10}$
and one sees that ours is the non-Abelian
generalization of that physics.  In particular $m$ is recognized as the gauge
invariant Debye screening length.

At the present time the non-Abelian equations  are under further
investigation.  It has been    possible to find local expressions for the
current in the static case$^{9,11}$
and in the position-independent case.$^{11}$
$$
\eqalign{
{m^2 \over 2} j^\mu_{\rm induced}
&= (-m^2 A_0, \pmb{0}) \cr
{m^2 \over 2} j^\mu_{\rm induced}
&= (0, -{1\over3} m^2 {\bf A}) \cr}
{}~~~~\eqalign{\vphantom{1\over1}&\hbox{(static)}\cr
           \vphantom{1\over1}&\hbox{(position-independent)}\cr}
$$
The non-Abelian Kubo equation may then be solved, but the physical relevance
of the solutions is unclear.  In particular the static solutions are not
solitons, since their energy is infinite.

A much more interesting result is due to Blaizot and Iancu.$^{12}$
They abstract from the Silin solution the plane-wave
{\it Ansatz\/} $A_\mu(x) = A_\mu(x\cdot p)$
where $p$ is a constant 4-vector and they determine
explicitly the induced current associated with non-Abelian plane waves.   In
terms of the above, this corresponds to
$$
a_\pm =
{Q_\pm \cdot p \over Q_\mp \cdot p} \, A_\mp
$$
%% ~~
%% \vcenter{
%% \hbox{\vrule%
%% \vbox{\hrule\vskip3pt\hbox{\hskip3pt(something)\hskip3pt}\vskip3pt\hrule}%
%% \vrule}}

The physics of all these solutions, as well as of other, still undiscovered
ones, remains to be elucidated, and I invite any of you to join in this
interesting task.

\goodbreak\bigskip
\centerline{\bfmone REFERENCES}
\nobreak
{\parskip=10truept
\raggedright

\item{1.~}
R.~Pisarski,
``How to compute scattering amplitudes in hot gauge theories'',
\hbox{{\it Physica\/}~{\bf A158}}, 246 (1989).

\item{2.~}
E.~Braaten and R.~Pisarski,
``Soft amplitudes in hot gauge theory: a general analysis'',
{\it Nucl.~Phys.\/}~{\bf B337}, 569 (1990);
J.~Frenkel and J.~C.~Taylor,
``High-temperature limit of thermal QCD'',
{\it Nucl.~Phys.\/}~{\bf B334}, 199 (1990).

\item{3.~}
J.~C.~Taylor and S.~M.~Wong,
``The effective action of hard thermal loops in QCD'',
{\it Nucl.~Phys.\/}~{\bf B346}, 115 (1990).

\item{4.~}
R.~Jackiw,
``Gauge theories in three dimensions
($\approx$ at finite temperature)''
in {\it Gauge Theories of the Eighties\/},
R.~Ratio and J.~Lindfors, eds.
Lecture Notes in Physics {\bf 181}, 157 (1983)
(Springer, Berlin, 1983).

\item{5.~}
R.~Efraty and V.~P.~Nair,
``Action for the hot gluon plasma based on the Chern-Simons theory'',
{\it Phys.~Rev.~Lett.\/}~{\bf 68}, 2891 (1992);
``Chern-Simons theory and the quark-gluon plasma'',
{\it Phys.~Rev.~D\/} {\bf 47}, 5601 (1993).

\item{6.~}
D.~Gonzales and A.~Redlich,
``A gauge invariant action for 2+1 dimensional topological Yang-Mills theory'',
{\it Ann.~Phys.\/} (NY) {\bf 169}, 104 (1986);
G.~Dunne, R.~Jackiw and C.~Trugenberger,
``Chern-Simons theory in the Schr\"odinger representation'',
{\it Ann.~Phys.\/} (NY) {\bf 149}, 197 (1989).

\item{7.~}
R.~Jackiw and V.~P.~Nair,
``High temperature response function and the non-Abelian Kubo formula'',
{\it Phys.~Rev.~D\/}~{\bf 48}. 4991 (1993).

\item{8.~}
J.-P.~Blaizot and E.~Iancu,
``Kinetic Equations for long-wavelength excitation of quark-gluon plasma'',
{\it Phys.~Rev.~Lett.\/} {\bf 70}, 3376 (1993);
``Soft collective excitations in hot gauge theories'',
{\it Nucl.~Phys.\/}~{\bf B} (in press).

\item{9.~}
R.~Jackiw, Q.~Liu, and C.~Lucchesi,
``Hard thermal loops,
static response and composite effective action'',
MIT preprint CTP \#2261 (November 1993).

\item{10.~}
E.~M.~Lifshitz and L.~P.~Pitaevskii,
{\it Physical Kinetics\/} (Pergamon, Oxford, 1981).

\item{11.~}
J.-P.~Blaizot and E.~Iancu,
``Non-Abelian excitation of the quark-gluon plasma''.
Saclay preprint T94/002.

\item{12.~}
J.-P.~Blaizot and E.~Iancu,
``Non-Abelian~plane~waves~in~the~quark-gluon~plasma'', Saclay preprint T94/013.

}

\bye